# Magnetic diffusion in Solar atmosphere produces measurable electric fields

Tetsu Anan[1]*, Roberto Casini[2], Han Uitenbroek[3], Thomas A. Schad[1], Hector Socas-Navarro[4,5], Kiyoshi Ichimoto[6], Sarah A. Jaeggli[1], Sanjiv K. Tiwari[7,8], Jeffrey W. Reep[9], Yukio Katsukawa[10], Ayumi Asai[11], Jiong Qiu[12], Kevin P. Reardon[3,13], Alexandra Tritschler[3], Friedrich Wöger[3], Thomas R. Rimmele[3]

[1]National Solar Observatory, Makawao, Hawaii, 96768, USA
[2]High Altitude Observatory, National Center for Atmospheric Research, Boulder, Colorado, 80301, USA
[3]National Solar Observatory, Boulder, Colorado, 80303, USA
[4]Institute de Astrofísica de Canarias, La Laguna, Tenerife, Spain
[5]Departamento de Astrofísica, Universidad de La Laguna, 38205, Tenerife, Spain
[6]College of Science and Engineering, Ritsumeikan University, Kusatsu, Shiga, 525-8577, Japan
[7]Lockheed Martin Solar and Astrophysics Laboratory, Palo Alto, California, 94304, USA
[8]Bay Area Environmental Research Institute, NASA Research Park, Moffett Field, California, 94035, USA
[9]Institute for Astronomy, University of Hawai`i at Mānoa, Pukalani, HI, 96768, USA
[10]National Astronomical Observatory of Japan, Mitaka, Tokyo 181-8588, Japan
[11]Astronomical Observatory, Kyoto University, Sakyo, Kyoto 606-8502, Japan
[12]Department of Physics, Montana State University, Bozeman, MT, 59717, USA
[13]Astrophysical and Planetary Sciences Department, University of Colorado, Boulder, CO, USA

Correspondence and requests for materials should be addressed to T.A. (email: tanan@nso.edu)

**The efficient release of magnetic energy in astrophysical plasmas, such as during solar flares, can in principle be achieved through magnetic diffusion, at a rate determined by the associated electric field. However, attempts at measuring electric fields in the solar atmosphere are scarce, and none exist for sites where the magnetic energy is presumably released. Here, we present observations of an energetic event using the National Science Foundation's Daniel K. Inouye Solar Telescope, where we detect the polarization signature of electric fields associated with magnetic diffusion. We measure the linear and circular polarization across the hydrogen Hε Balmer line at 397 nm at the site of a brightening event in the solar chromosphere. Our spectro-polarimetric modeling demonstrates that the observed polarization signals can only be explained by the presence of electric fields,**

**providing conclusive evidence of magnetic diffusion, and opening a new window for the quantitative study of this mechanism in space plasmas.**

# Introduction

Magnetic reconnection is a ubiquitous process in laboratory and astrophysical plasmas wherein anti-parallel magnetic field components interact diffusively, resulting in a change of the field topology as well as energy conversion to internal and kinetic energies of the embedded plasma[1-3]. Reconnection can cause various kinds of transient phenomena, such as $\gamma$ - ray bursts in distant galaxies, eruptive energy release in the solar atmosphere, substorms in the Earth's magnetosphere, and disruption of fusion plasmas in the laboratory[4]. However, for solar and stellar plasmas, no conclusive observational evidence of magnetic diffusion, which allows the magnetic field lines to reconnect, has been reported through remote sensing.

One expected ingredient of magnetic diffusion is the manifestation of electric fields in finite-conductivity plasma. This expectation arises from Faraday's law,

$$\frac{\partial \mathbf{B}}{\partial t} = -\nabla \times \mathbf{E}, \quad (1)$$

where **B** is the magnetic field and **E** is the electric field. The curl of the electric field vector can be reduced to a diffusion of the magnetic field,

$$\frac{\partial \mathbf{B}}{\partial t} = \frac{1}{\sigma\mu} \Delta \mathbf{B}, \quad (2)$$

using Ohm's law, $\mathbf{J} = \sigma\mathbf{E}$, and Ampére's law assuming the displacement current is negligibly small, $\nabla \times \mathbf{B} = \mu\mathbf{J}$, where $\sigma$ is the conductivity, and $\mu$ is the permeability. However, the possibility of measuring electric fields in space plasmas has been given little attention, under the presumption that the electric fields dissipate quickly or exist on an unresolvable spatial scale due to the high electrical conductivity of the solar plasma.

These electric fields can, in principle, be observationally inferred using spectro-polarimetry. There are two different ways in which the presence of electric fields in plasma can affect the polarization of spectral lines from hydrogenic atoms. One is the well-known energy splitting of the atomic level due to the linear Stark effect[5,6]. This is expected to be the dominant effect in regions of the solar atmosphere that can harbor electric fields while being characterized by plasma conditions typical of local thermodynamic equilibrium (e.g., in the photosphere)[7-11]. The second effect is the modification of the scattering polarization induced by the optical pumping of the atom by the anisotropic radiation from the underlying atmosphere. Similar to the Hanle effect and the alignment-to-orientation (A-O) conversion mechanism[12] that have been observed in various lines formed in magnetized chromospheric plasmas (e.g., the He I D$_3$ line in prominences[13]), for hydrogenic lines an analogous effect is established in the presence of an electric field[14].

In particular, the A-O mechanism is directly responsible for the appearance of a symmetric Stokes-$V$ profile. For hydrogen lines that are formed in a magnetic-only environment, the A-O mechanism requires magnetic field strengths large enough to mix the fine structure levels of the transition terms (typically, a few tenths of a tesla, T). For such large field strengths, the circular polarization is largely dominated by the Zeeman effect, so the resulting net circular polarization cannot qualitatively modify the antisymmetric appearance of the Stokes-$V$ profile. In contrast, in the presence of an electric field, the A-O mechanism sets in at energy separations of the order of the Lamb shifts (approximately one order of magnitude smaller than the fine-structure separation) between adjacent atomic terms "mixed" by the electric field[15]. This is the reason why a net-circular-polarization-dominated, symmetric Stokes-$V$ profile can be produced in hydrogen lines even in the presence of relatively small magnetic strengths, when an electric field is also present[14,16].

For these reasons, polarimetric measurements of various spectral lines of neutral hydrogen have been conducted in the attempt to diagnose electric fields in the solar atmosphere. During the 1980s and early 1990s, those measurements targeted the linear polarization of H I lines from highly excited levels, using the differential Stark broadening of the $I + Q$ and $I - Q$ profiles as a measure of the electric field strength[7]. These measurements were performed for the apex of a loop formed during the late phase of a solar flare[8], the middle of a "surge" [9,10], a type of solar jet with chromospheric temperatures, i.e., approximately $10^4$ K. Solar prominences were also observed by Foukal and collaborators[9-11], but polarization signals of the electric field were not conclusively detected. In 2005, a symmetric profile in the circular polarization of Hα, which could not be explained by the magnetic A-O mechanism, was measured in a prominence[17]. The authors proposed the presence of electric fields as a possible explanation, although an estimation of the electric strength could not be provided at that time.

In this work, we present measurements of polarization signals that can only be explained by the presence of an electric field at a null point of the magnetic field in the lower chromosphere. Because the signal is detected at the site of a brightening event that has been interpreted as a result from magnetic reconnection, we conclude that this observation provides conclusive evidence of magnetic reconnection triggering the energetic event. Furthermore, the spatial scale of the electric-field signature allows us to constrain reconnection models working for the active phenomenon.

# Results

**Observations**

New possibilities for detecting electric fields on the solar surface are now available using the world's largest optical solar telescope, the US National Science Foundation's Daniel K. Inouye Solar Telescope (DKIST[18]) at Haleakalā on the Hawaiian island of Maui, which began principal-investigator-led science observations in December 2021. The first object successfully observed was an active region named NOAA 12955 at the heliocentric coordinates of N14° E44° on 23 February 2022 (Fig. 1). One of the facility instruments of DKIST, the Visible Spectro-Polarimeter (ViSP[19]), measured the full state of polarization in the spectral regions of Ca II H 396.9 nm, the Fe I doublet 630.2 nm, and Ca II 854.2 nm (see, methods subsection ViSP observations), together with blue continuum imaging (450.3 nm) provided by another DKIST facility instrument, the Visible Broadband Imager (VBI[20]). The spectral region of the Ca II H chromospheric line also contains the Hε Balmer line of neutral hydrogen at 397.0 nm, typically formed in the lower chromosphere[21].

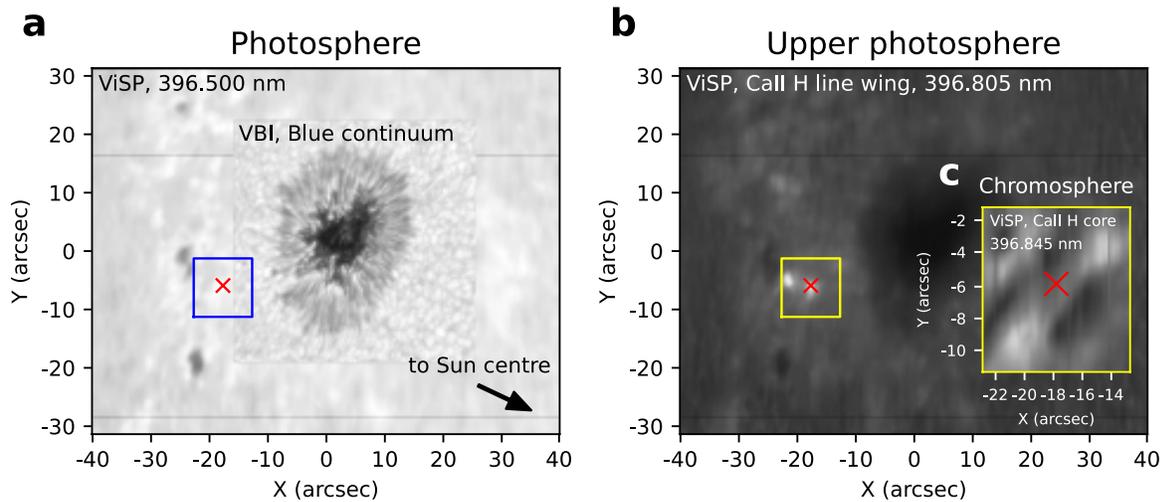

Figure 1 | **Active region observed by the NSF's Daniel K. Inouye Solar Telescope (DKIST[18]) on 23 February 2022. a** Solar photosphere shown by a Visible-Spectro-Polarimeter (ViSP[19]) intensity map at 396.5 nm and an inset Visible-Broadband Imager (VBI[20]) blue continuum image at 450.3 nm. The spectrograph slit of the ViSP scanned over the active region in the +X-direction to produce this map. The two horizontal black lines in the ViSP map are images of fiducials at the slit used for alignment purposes. The black arrow points towards the solar disc centre. **b** Upper photosphere and lower chromosphere shown by a ViSP intensity map in the Ca II H line wing at −0.05 nm from the line core, in which the line has a line wing emission. In general, Ellerman bombs are bright in this wavelength. Within the yellow square, two Ellerman bombs were observed. The red cross points to the location of the Ellerman bomb and the place where the Stokes profiles shown in Fig. 3 and 4 were measured. The other is the bright feature at (X,Y) = (−22 arcsec, −5 arcsec). In this study, we present results for the former one. The blue and yellow squares mark the region enlarged in Fig. 2. **c** Chromosphere within the square is shown at the line core of Ca II H.

We observed an Ellerman bomb between the sunspot and pores in the active region, marked by the red cross in Figs. 1 and 2 (see, methods subsection The Ellerman bomb observed with SDO/AIA, and supplementary notes subsections Stokes-*I* profile in the Hε line of the Ellerman bomb). Ellerman bombs, discovered[22] in 1909, are phenomena that are distinctly bright in the wings of spectral lines, e.g., Hα and Ca II H, but typically not at their line cores[23] (Fig. 1b). They have been interpreted as sites of magnetic reconnection releasing magnetic energy and making the lower chromosphere dense and hot[24-27]. The observed brightening occurred near the border between two regions where magnetic fields were directed respectively inward and outward of the solar surface (Fig. 2d). In addition, it seemed to be accompanied by a downflow, while most plasma moved upward in the surrounding regions (Fig. 2b), although it is not clear whether the downflow was one of the reconnection bi-directional outflows. Thus, magnetic reconnection was likely to occur in this antiparallel magnetic field configuration, leading to the observed brightening (see, methods subsections The 180° azimuthal ambiguity, and supplementary notes subsection Line-of-sight magnetic field in the photosphere).

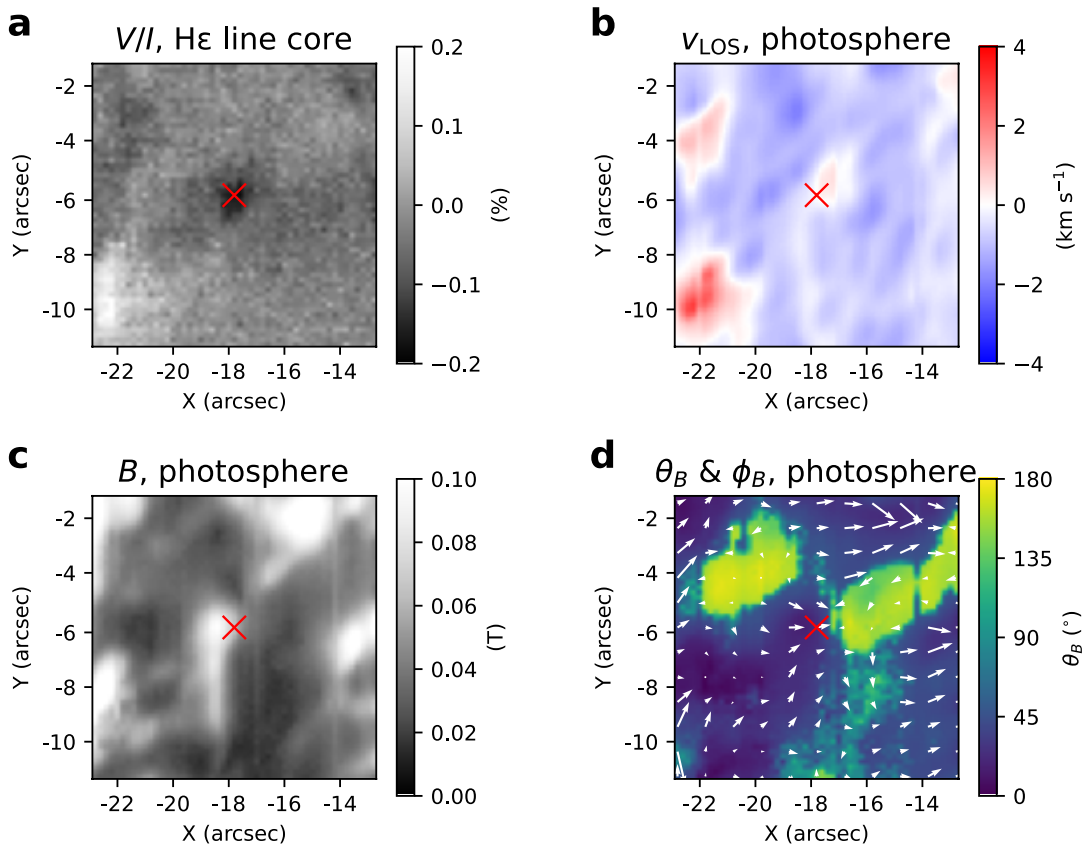

Figure 2 | **Measured quantities of the Ellerman bomb.** The red cross points to the Ellerman bomb and the place where Stokes profiles shown in Figs. 3 and 4 were measured. **a** Stokes-*V/I* map at the line centre of the Balmer line of neutral hydrogen, Hε, at 397.0 nm. Idealized Stokes profiles generated by the Zeeman effect should be zero in this map, although Stokes profiles can be complicated at a high spatial resolution. Non-zero values indicate that the Stokes-*V* profile

was either deformed by gradients of velocity and magnetic fields along the line of sight[28,29], or another mechanism created the observed broadband circular polarization, e.g., the alignment-to-orientation (A-O) conversion mechanism induced by the Stark effect[14,15]. **b** Line-of-sight component of velocity, $v_{\text{LOS}}$, in the photosphere derived from the Doppler shift of the Fe I 630.15 nm line through a single Gaussian fitting. Plasma moving downward and upward relative to the plasma at the red cross is marked in red and blue, respectively. **c** Magnetic field strengths, $B$, in the photosphere inferred from the Fe I 630.15 nm line through the Weak Field Approximation method[12]. **d** Inclination angles from the solar local vertical of the magnetic field, $\theta_B$. The arrows show the azimuthal direction of the magnetic field, $\phi_B$. Their lengths are scaled by the strength of the horizontal component of the magnetic fields. The 180° azimuthal ambiguity in a line-of-sight reference frame[12] was resolved (see, methods subsection The 180° azimuthal ambiguity). The line-of-sight component of the magnetic field in the photosphere is also presented in the supplementary information (see, subsection Line-of-sight magnetic field in the photosphere).

**Electric-field diagnosis**

Associated with the Ellerman bomb, we detected a clear signal of net circular polarization in Hε (Fig. 2a), with a Stokes-$V$ profile that was symmetric with respect to line centre (Figs. 3d and 4c, see also, supplementary notes subsection Another Ellerman bomb with electric field signals). After considering possible alternative mechanisms that can produce net circular polarization via the Zeeman effect, i.e., due to the presence of strong fields and velocity gradients in the line formation region[28,29], we were unable to explain the observed circular polarization signature. The action of an ambient electric field has been proposed as a possible mechanism to explain such a symmetric Stokes-$V$ profile in hydrogen lines[14,17]. In this work, we provide observational evidence and modeling results to support this explanation.

We synthesized the Stokes profiles of Hε taking into account only magnetic effects in order to demonstrate that an external electric field is required to reproduce the observed profiles. First, the Stokes profiles of four spectral lines (not including Hε), Ca II H 397 nm, Fe I 630.15 nm, Fe I 630.25 nm, and Ca II 854 nm, which are formed in the photosphere and the chromosphere, were fitted starting from 10,000 different atmospheric model guesses using the state-of-the-art inversion code, Departure coefficient aided Stokes Inversion based on Response functions (DeSIRe[30]) (see, methods subsection Inversion with DeSIRe). We selected the 50 best fits, and additionally synthesized the Stokes profiles of Hε using the corresponding inferred atmospheres through DeSIRe (red lines in Fig. 3). The results show that the Stokes $I$ and $V$ profiles of all four lines, i.e., Ca II H 397 nm, Fe I 630.15 nm, Fe I 630.25 nm, and Ca II 854 nm, are fitted well, apart from a blue bump in Stokes $I$ of Ca II 854 nm and Stokes $V$ at the corresponding wavelength (Fig 3g and h). In contrast, the synthesized Stokes-$V$ profiles of Hε are not symmetric with respect to the line centre as shown by the observations, demonstrating that the observed Hε Stokes profiles cannot be qualitatively explained using the magnetic fields in the stratified atmospheres inferred from the other lines. If we fitted only for Ca II 854 nm, we could reproduce the observed Stokes-*I* and -*V* profiles including the blue bump, and confirmed that the observed Stokes-*V/I* profile in the Hε line could not be explained using only the magnetic fields in the stratified atmospheres (see, methods subsection DeSIRe inversion only for Ca II 854 nm). This conclusion is also confirmed using a 3D radiative magnetohydrodynamics (MHD) numerical model (see, methods subsection Hε Stokes profiles of Ellerman bombs from a MHD simulation)

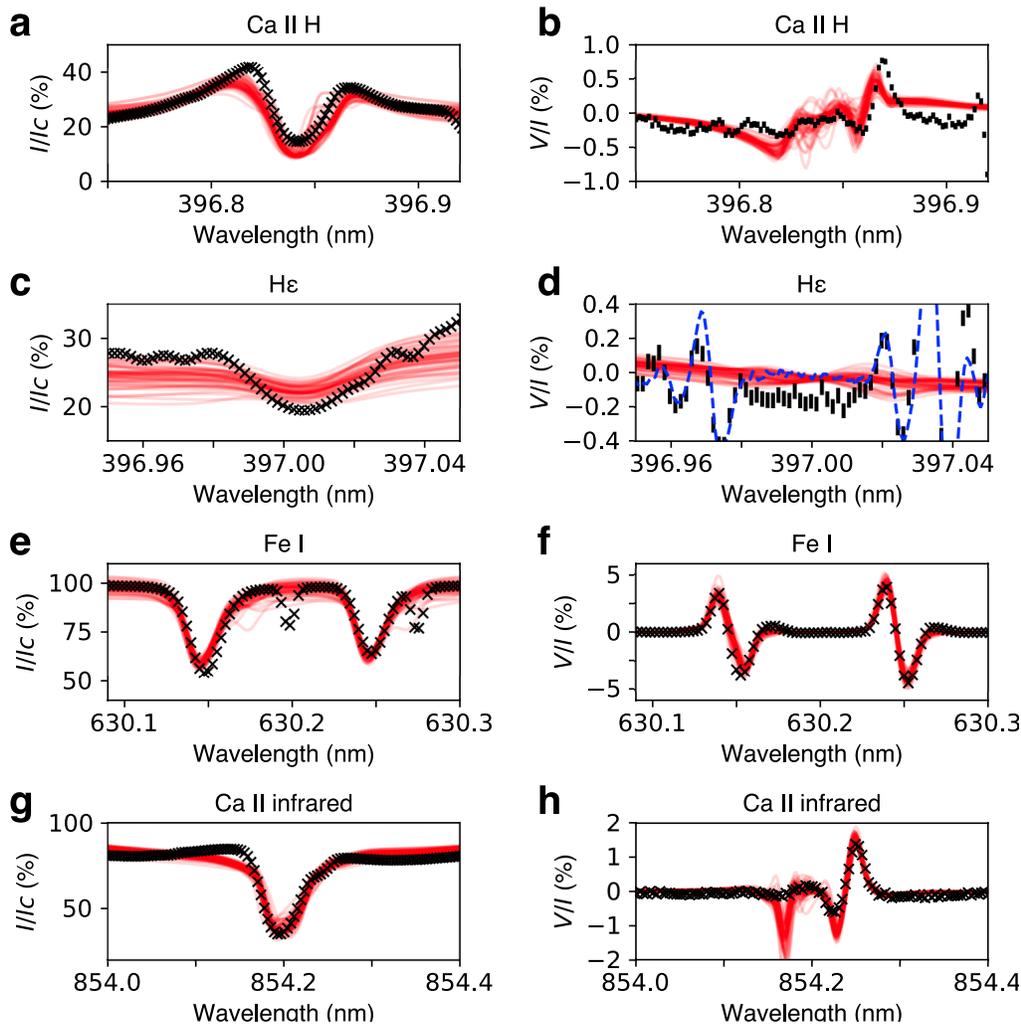

Figure 3 | **Stokes-*I* and *V/I* profiles of the Ellerman bomb** at the location marked by the red cross in Figs. 1 and 2. The left column shows Stokes-*I* profiles normalized by the intensity at the continuum, $I_c$, in the Ca II H (**a**), Hε (**c**), Fe I 630.15 nm and Fe I 630.25 nm (**e**), and Ca II 854 nm lines (**g**), and the right column shows Stokes-*V/I* profiles in the Ca II H (**b**), Hε (**d**), Fe I 630.15 nm and Fe I 630.25 nm (**f**), and Ca II 854 nm lines (**h**). Black crosses and bars are measured data points. Error bars are shown as the standard deviation in the continuum only for

Stokes-*V/I* in the Ca II H and Hε, because they are too small to see in the other plots. Red lines show synthesized profiles from the best 50 fitting results of Departure coefficient aided Stokes Inversion based on Response functions (DeSIRe[30]) inversions, which were run 10,000 times while varying the initial atmospheric model (see, methods subsection Inversion with DeSIRe). A blue dashed curve in panel d shows an average profile of those pixels that have similar Stokes-*V/I* signals to the Ellerman bomb in weak photospheric lines, i.e., Fe I at 396.96 nm, Cr I at 396.97 nm, Fe I at 397.02 nm, and Ni I at 397.04 nm, from quiet regions around the Ellerman bomb. The narrow spectral lines in 630.20 and 630.28 nm are telluric lines of $O_2$. Source data are provided as a Source Data file.

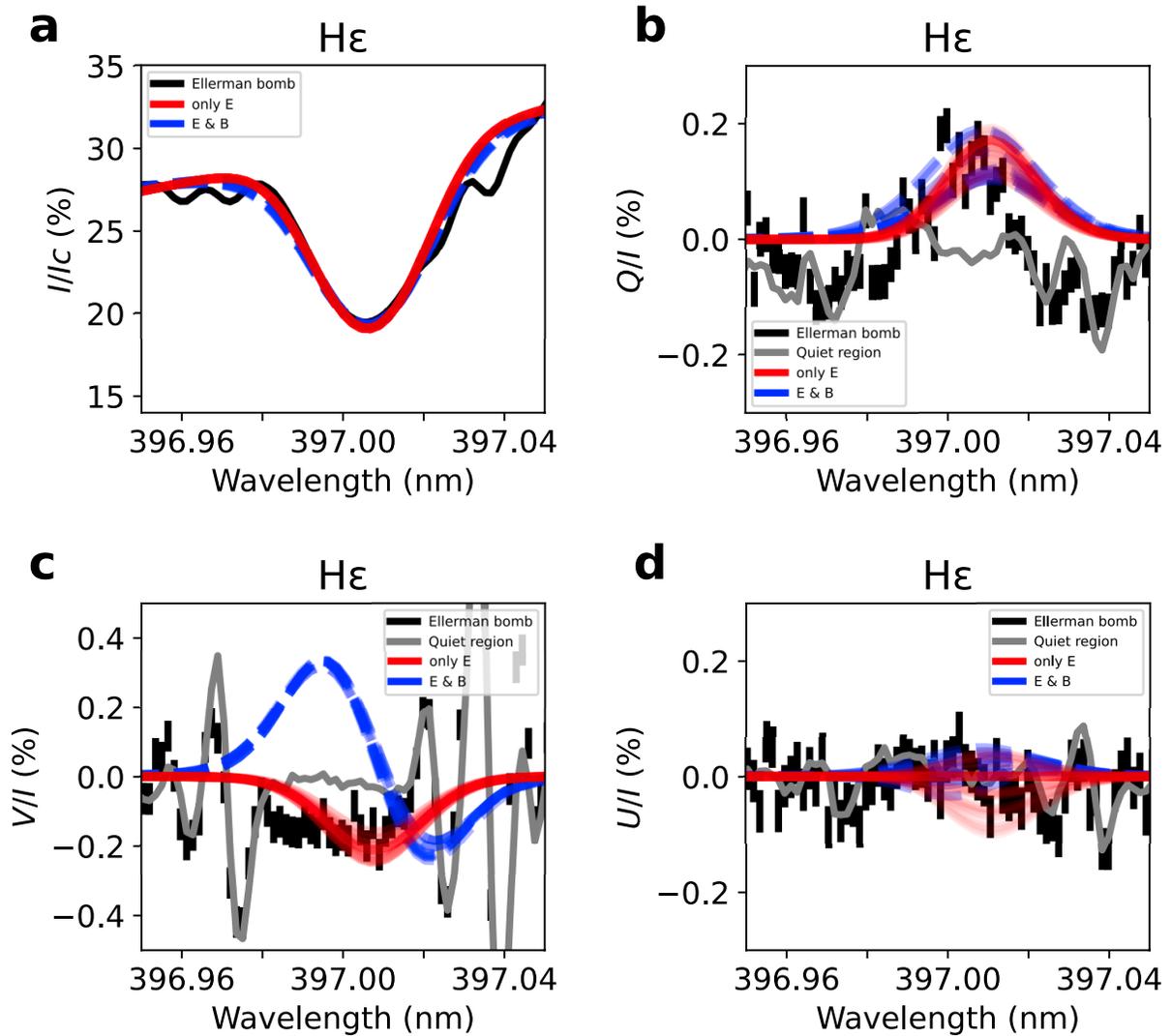

Figure 4| **Hε synthesized with electric fields. a** Stokes-*I*, **b** Stokes-*Q/I*, **c** Stokes-*V/I*, and **d** Stokes-*U/I* profiles in the Balmer line of neutral hydrogen, Hε, at 397.0 nm. Black curves and bars are measured data points at the location marked by the red cross in Fig. 1 and 2. Error bars

are shown as the standard deviation in the continuum only for Stokes-$Q/I$, $U/I$, and $V/I$, because they are too small for Stokes-$I$. The gray curves were derived from quiet regions in the same way as the derivation of the blue dashed curve in Fig. 3(d) to show the Stokes signals of the weak photospheric lines. The red solid lines are profiles of Hε synthesized in the presence of an electric field and no magnetic fields. We calculated them while varying the three components of the electric field vector by 10 V m$^{-1}$ in strength and 5° in inclination and azimuth angles. Then, we selected some that reproduce the amplitudes of the measured profiles. The blue dashed lines are similar to that of the red lines, but with the addition of a photospheric magnetic field as inferred through the Weak Field Approximation[12] using the Fe I 630.15 nm line (Fig. 2). Source data are provided as a Source Data file.

Instead, we found that the observed Stokes-$V/I$ profile of Hε can be reproduced if we assume the presence of an ambient electric field. This is possible because of the net circular polarization produced by the A-O mechanism induced by the Stark effect[14,15]. The blue dashed lines in Fig. 4 show Stokes profiles calculated with various electric fields in addition to a magnetic field inferred through the Weak Field Approximation[12] using the Fe I 630.15 nm line (Fig. 2) taking into account the A-O mechanism induced by the Paschen-Back effect[31,32] and the Stark effect[14] (see, methods subsection Calculation of hydrogen Stokes profiles with electric fields). However, the synthesized Stokes profiles exhibit significant deviations from the observed profiles, particularly in the blue lobe of the Stokes-$V$ profile. In contrast, assuming that the line was formed at the location of a magnetic null-point at the time of observation, we were able to identify possible electric field configurations that can reproduce the observed Stokes profiles (red solid lines in Fig. 4). The inferred electric fields have two degenerate solutions pertaining to mostly vertical or horizontal geometries with respect to the local solar vertical. The nearly horizontal solution has a strength of $110 \pm 10$ V m$^{-1}$, with an inclination from the solar vertical of $75 \pm 5°$, and an azimuth angle of $-5 \pm 20°$ measured from the radial direction from Sun centre to the observed region (i.e., roughly antiparallel to the black arrow in Fig. 1a). The nearly vertical one has a strength of $107.5 \pm 7.5$ V m$^{-1}$ and an inclination of $177.5 \pm 2.5°$ (because the field is practically vertical, the azimuth is poorly defined). Because magnetic reconnection in the lower solar atmosphere is expected to occur at the boundary between oppositely directed vertical magnetic fields (Fig. 2d), the electric current in this situation, calculated as $\frac{1}{\mu} \nabla \times \mathbf{B}$, should be horizontal to the solar surface. It is implied that the horizontal solution is the most likely, assuming the electric field was parallel to the electric current (Fig. 5).

# Discussion

We conclude on the basis of these results that a nearly horizontal electric field existed at a null point of the magnetic field during the observation. This conclusion agrees well with the magnetic reconnection in the antiparallel magnetic field structure, which was expected for this Ellerman bomb (Fig. 2). Assuming a typical Alfvén velocity of 100 km s$^{-1}$ and a strength of the reconnecting magnetic fields of 0.05 T, the reconnection rate defined as the electric field normalized by the product of the magnetic field and the Alfvén velocity would be 0.02, which is consistent with the rates estimated through other methods[33,34].

The spatial scale of the diffusion of the magnetic field has been thought to be smaller than 1 km[35]. However, the measured signals of the negative Stokes-*V/I* at the line centre of Hε filled a region up to 1 arcsecond in size, which is approximately 700 km on the solar surface (Fig. 2a). If this is the true size of the reconnection region, it may indicate that a large number of unresolved small diffusion regions contributed to the signal coming from the observed area. Such diffusion-region "assemblies" have been proposed within turbulent reconnection[36] and plasmoid-induced reconnection[37]. Because the aspect ratio of the diffusion region should be greater than $10^4$ to produce the plasmoids[38-40], our observations support the turbulent reconnection model (Fig. 5).

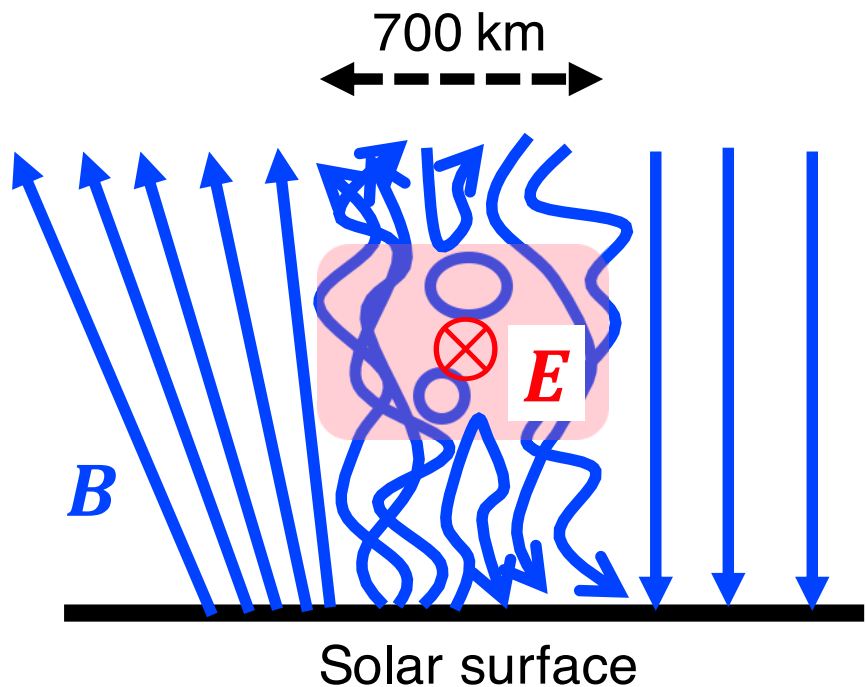

Figure 5| **Magnetic and electric field configuration associated with the Ellerman bomb.** The blue curves are magnetic field lines. The red region marks where the electric field signal was supposedly measured. The symbol ⊗ indicates that the electric field is directed into the page.

# Methods

**ViSP observations**

We measured the full state of polarization in three windows of the solar spectrum, 396.4 – 397.2 nm, 629.5 – 630.8 nm, and 853.2 – 855.0 nm on 23 February 2022 with ViSP using its three spectral arms. The physical detector pixel size is about 6.5 µm. Spectral samplings were 0.77 pm pixel$^{-1}$ for 397 nm, 1.3 pm pixel$^{-1}$ for 630 nm, and 1.9 pm pixel$^{-1}$ for 854 nm. Spatial samplings along the slit for 397, 630, and 854 nm were 0.0245, 0.0298 and 0.0194 arcsec pixel$^{-1}$, respectively. We binned the data in the spatial direction by 8 for 397 nm, 6 for 630 nm, and 10 for 854 nm to increase the sensitivity. The Stokes profiles in Fig. 3 and 4 were measured at 21:23 UT. They come from one pixel marked by the red cross in Fig. 1 and 2 after this spatial binning performed.

The spectrograph slit of the ViSP used for these observations had a width of 0.2142 arcsec, and it was scanned twice over the active region for a total of 400 steps each. The scanning width and duration were 86 arcsec and 106 minutes, respectively. The Ellerman bomb was observed in the second scan. For each slit step, the signal was acquired with an exposure time of 4 ms for 630 and 854 nm, and 20 ms for 397 nm. It was integrated over 50 modulation cycles of 10 states each. The step cadence was 16 seconds, which is significantly smaller than the typical lifetime of Ellerman bombs[26,41]. The slit was oriented perpendicular to the local horizon, so that the atmospheric refraction would shift the image along the slit direction[42].

**The Ellerman bomb observed with SDO/AIA**

Ellerman bombs often are bright in extreme ultraviolet (EUV) wavelengths, 160 nm and 170 nm[43]. Supplementary Fig. 1 shows EUV images of the observed Ellerman bomb in the wavelengths taken with Atmospheric Imaging Assembly (AIA[44]) onboard Solar Dynamics Observatory (SDO[45]), which continuously observes the full disk of the sun every 12 seconds. The data was taken at 21:23 UT on 23 February 2022. The Ellerman bomb was bright in these channels. This fact supports our interpretation that the observed brightening was an actual Ellerman bomb.

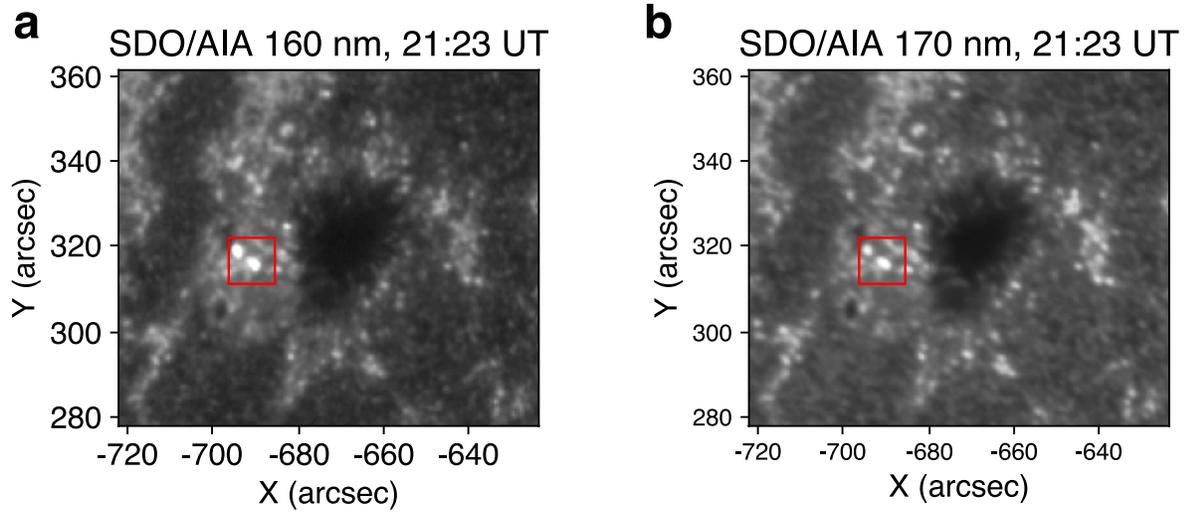

Supplementary Fig. 1 | **Ellerman-bomb images taken by Atmospheric Imaging Assembly (AIA[1]) onboard Solar Dynamics Observatory (SDO[2])** in 160 nm (**a**) and 170 nm (**b**). The observed Ellerman bomb is at the center of the red square. Level 1.5 data product was downloaded (see, Data availability) and plotted.

**The 180° azimuthal ambiguity**

In principle, the inferred azimuth of the magnetic field vector in the line-of-sight reference frame has a 180° ambiguity due to the Zeeman effect. We resolved the ambiguity for the magnetic field inclination and azimuth in the solar reference frame, shown in Fig. 2(d), by comparing the two candidate solutions with that found in the associated Space weather HMI Active Region Patches (SHARPs) data set[46]. A minimum energy method has been used to resolve the ambiguity in the SHARP data[47,48]. In addition, we selected the other solution, in which the magnetic field directs toward the sun center, for a magnetic patch located at $(X, Y) = (-16", -5")$ in Fig. 2(d), while the SHARP solution indicates the magnetic field was almost on the horizontal plane of the solar surface. Because the magnetic field was reconfigured by magnetic reconnection at the Ellerman bomb, it would be possible that the local magnetic field could be different from that in the SHARP data. Although the magnetic field in either solution can reconnect to the magnetic field at the Ellerman bomb, we selected the one that can produce stronger electric current.

**Inversion with DeSIRe**

DeSIRe[30] can simultaneously solve the radiative transfer equations for multiple lines of multiple atoms in a stratified and magnetized atmosphere under non-local-thermodynamic-equilibrium conditions, taking into account the Zeeman effect. We applied DeSIRe with two components to the Stokes profiles of the Ca II H 396.85 nm, Fe I 630.15 nm, Fe I 630.25 nm, and Ca II 854.21 nm lines simultaneously to infer physical quantities of the Ellerman bomb. In order to estimate meaningful ranges for the values of the atmospheric parameters, the inversion was performed many times using different initial guesses for some of those parameters, in the way described below.

We started by fitting only the Stokes profiles of the Fe I 630 nm lines using the Harvard Smithsonian Reference Atmosphere (HSRA[49]) as the initial guess. Because the observed Stokes profiles in the Ca II lines at the Ellerman bomb depart too much from those synthesized ones using the inferred atmosphere to be fitted, we manually modified the atmosphere and added HSRA atmosphere as the second component with certain magnetic field vectors and line-of-sight velocities. Then, we used the two modified atmospheres as the initial guess for the inversion of all the other lines (with the exception of Hε) varying a critical set of atmospheric parameters: temperature, line-of-sight velocity, and magnetic field strength in the atmosphere's stratification. This exercise produced the set of 10,000 inversion solutions out of which we preserved only the best 50 fits.

**DeSIRe inversion only for Ca II 854 nm**

We could not reproduce a part of the Stokes profiles in the Ca II 854 nm line around 854.15 nm, when we fitted the four spectral lines, i.e., Ca II H 397 nm, Fe I 630.15 nm, Fe I 630.25 nm, and Ca II 854 nm (Fig. 3). In order to confirm that the failure of the fitting does not change our conclusions, we applied the DeSIRe inversion only to the Ca II 854 nm profiles and synthesized the Stokes profiles of Hε using the inferred atmospheres (Supplementary Fig. 2). The Stokes-$I$ and most part of the Stokes-$V$ profile in the Ca II line was reproduced well, but not the observed Hε Stokes-$V/I$ profile. We confirmed on the basis of these fitting results that the observed Hε Stokes profiles cannot be explained using the inferred atmospheres, which can reproduce the Ca II 854 nm profiles and can be more varied than that inferred for the four spectral lines.

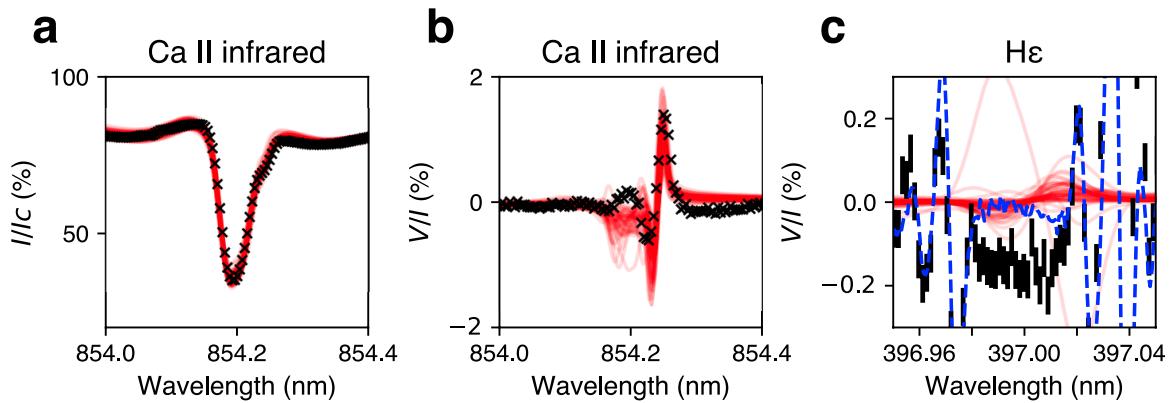

Supplementary Fig. 2| **Inversions applied only to the Ca II 854 nm line.** Stokes-$I$ and $V/I$ profiles of the Ellerman bomb, the same as Fig. 3, but only for the Ca II 854 nm and Hε lines. **a** Stokes-$I$ profiles in Ca II 854 nm normalized by the continuum intensity, $I_c$. **b** Stokes-$V/I$

profiles in Ca II 854 nm. **c** Stokes-$V/I$ profiles in Hε. The red lines show synthesized profiles from the best 50 fitting results of Departure coefficient aided Stokes Inversion based on Response functions (DeSIRe[3]) inversions that were applied only to the Ca II 854 nm line. Source data are provided as a Source Data file.

**Hε Stokes profiles of Ellerman bombs from a MHD simulation**

Based upon the inverse spectral methods above, we find that electric fields are necessary in order to reproduce the observed symmetric Hε Stokes-$V/I$ profile. At the same time, the need for electric fields can further be established using forward spectral synthesis methods through advanced numerical models.

We considered an atmospheric model calculated with the radiative magnetohydrodynamics (MHD) simulation code, Bifrost[50-52], where two examples of Ellerman bombs were produced. In particular, we considered the same Bifrost snapshot and regions as in Kawabata et al.[53], who synthesized Ca II 854 nm Stokes profiles of Ellerman bombs (green and yellow boxes in Fig. 2 of Kawabata et al. [53]). Using the Rybicky-Hummer non-local-thermodynamic-equilibrium radiative transfer code (RH[54]), we synthesized Hε Stokes profiles for those two regions, to test whether we could reproduce symmetric Stokes-$V$ profiles similar to those we report on. We note that the RH code is the forward radiative transfer solver included in DeSIRe, which only accounts for the magnetic Zeeman effect.

The only Stokes-$V/I$ profiles produced by this simulation, showing a symmetric shape similar to the ones we observed, have absolute amplitude less than 0.03 %, compared to the negative amplitude of $-0.2$ % of the observed profile (Fig. 3 and 4), and therefore completely below detection. Supplementary Fig. 3 (a) shows a histogram of the net circular polarization (NCP), defined as $\int \frac{V}{I} d\lambda$, of the Hε profiles synthesized from the Bifrost Ellerman-bomb models. While the NCP of the observed profiles cannot be easily determined because of the blended weak spectral lines, the red profiles in Fig. 4 that are able to closely reproduce the observed ones all have NCPs of $-0.06 \pm 0.01$ nm, and so we can assume this value as a proxy for the observed NCP. In contrast, the absolute values of the NCP calculated from the Bifrost Ellerman-bomb models are three orders of magnitude smaller.

Supplementary Fig. 3 (b) shows the Stokes-$V/I$ profiles with the largest absolute values of the NCP in the Bifrost Ellerman-bomb model. Not only these profiles all have NCP values more than three orders of magnitude smaller than the observed ones, but their shape also remains largely anti-symmetric. Therefore, we can conclude that gradients of the magnetic fields and velocity in the line formation region[28,29] cannot reproduce the observed profile. The A-O mechanism due to the presence of an electric field at the null point of the magnetic field becomes a necessary ingredient to reproduce the observations, as discussed in the main text.

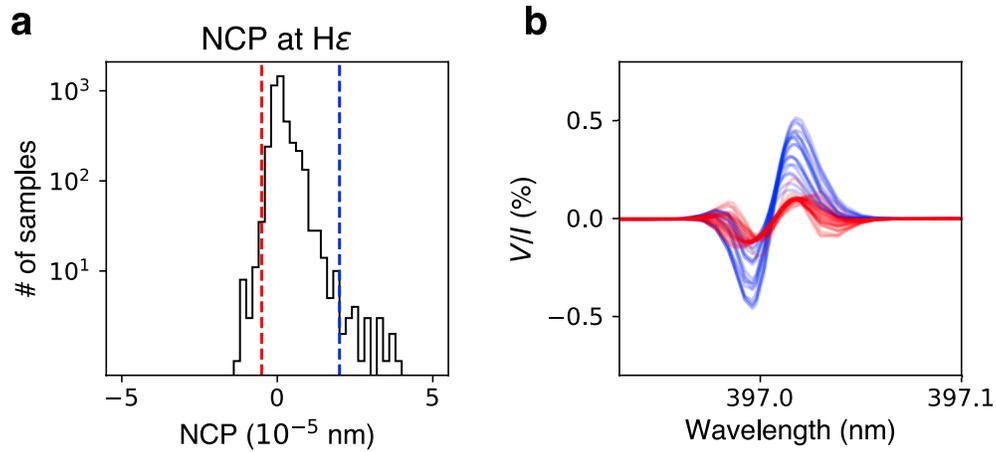

Supplementary Fig. 3| **Net Circular Polarization (NCP) of Hε Stokes-$V/I$ profiles synthesized from a Ellerman-bomb magnetohydrodynamics simulation result[4,5]. a** Histogram of NCP values derived from 4074 spectral profiles modeled in the vicinity of the Ellerman-bombs. **b** The blue and red curves show the Hε Stokes-$V/I$ profiles in the model having NCP, respectively, below the red threshold and above the blue threshold of panel **a**. Source data are provided as a Source Data file.

**Calculation of hydrogen Stokes profiles with electric fields**

The most established theoretical formalism for calculating polarized spectra of hydrogen lines in the presence of external magnetic and electric fields relies upon quantum density matrix theory[14]. In the limit of complete frequency redistribution, this formalism solves for the statistical equilibrium of quantum states of an ensemble of hydrogen atoms excited anisotropically by a prescribed external radiation field. The solution density matrix is used to calculate the local polarized absorptivity and emissivity properties of the plasma, and the emergent radiation field is subsequently calculated by integrating the standard polarized radiation transfer equation along the line-of-sight[12].

We use an established numerical code that implements this rather involved formalism[14,55] (see code availability statement). In this particular case, we calculate the Stokes profiles of Hε by adopting a simplified atomic model for hydrogen and an atmospheric model that consists of a single constant property slab of material illuminated from below by the lower solar atmosphere. The atomic model consists of the energy levels of principal quantum numbers, $n$ = 1, 2, 3, 7, with all the possible transitions among them, namely Lyman α, Lyman β, Lyman ζ, Hα, Hε, and Paschen δ. This model maintains the necessary complexity of the atomic system while minimizing the computation cost. Since radiation pumping drops fast with wavelength, the transitions between $n$ = 4 and 7, 5 and 7, and 6 and 7 that form spectral lines in the infrared would not significantly impact the current configuration.

We assumed that the radiation temperature of the underlying anisotropic continuum was 6000 K at all radiative transitions within the model. The degree of anisotropy for this pumping radiation field, which is assumed to be symmetric about the vertical axis, was derived for both Hε and Hα with RH[54] from an atmospheric model inferred by DeSIRe. For all other transitions we determined the anisotropy factor by adopting the center-to-limb variation model of Cox[56], with an ad-hoc extension for the Lyman series[57], and assuming a height of 1.4 Mm for the scattering atmosphere. Finally, we also added terms into the statistical equilibrium equations for describing simple term-to-term collisional transitions[14]. The rates for collisional de-excitation were set to $1.4 \times 10^6 \text{ sec}^{-1}$ for all levels. In addition, we assumed that the ground state was completely depolarized by collisions because of its very long lifetime.

**Data availability**

DKIST data used in the paper are publicly available on DKIST Data Center Archive at https://dkist.data.nso.edu/?proposalId=pid_1_50. The SDO data is publicly available at https://www.lmsal.com/solarsoft/ssw_service/queue/finished/ssw_service_231024_172634_35912.html. The Bifrost simulation data is also publicly available at https://sdc.uio.no/search/simulations?sim=qs024048_by3363. The time step we analyzed is 916. The Stokes spectra analyzed during the current study have been deposited in the Zenodo database[58], https://doi.org/10.5281/zenodo.13709863. The datasets generated during and/or analyzed during the current study are available from the corresponding author upon request. Source data are provided with this paper.

**Code availability**

The code used to make the observed Stokes profiles shown in Fig. 3 and 4 from data provided at the DKIST Data Center Archive and the code used to apply the DeSIRe inversion to the profiles are publicly available on the GitHub repository[59], https://doi.org/10.5281/zenodo.13621770. The Stokes inversion code, DeSIRe, is publicly available in the GitHub repository, https://github.com/BasilioRuiz. The code that was used to calculate the Stokes profiles of Hε in the presence of electric fields is not publicly available, but can be shared by the author (Dr. Roberto Casini; casini@ucar.edu) for the purpose of reproducibility tests or collaborative research relying on the scattering polarization of hydrogen lines as a diagnostic of magnetic and electric fields in astrophysical plasmas.

**References**

1. Sweet, P. A. The Neutral Point Theory of Solar Flares. *IAU Symposium* **6**, 123-134 (1958)

2. Dungey, J. W. The Neutral Point Discharge Theory of Solar Flare. A Reply to Cowling's Criticism. *IAU Symposium* **6**, 135-140 (1958)
3. Ji, H. et al. Magnetic reconnection in the era of exascale computing and multiscale experiments. *Nature Reviews Physics* **4**, 263-282 (2022)
4. Zweibel, E. G., Yamada, M. Magnetic Reconnection in Astrophysical and Laboratory Plasmas. *Annual Review of Astronomy & Astrophysics.* **47**, 291-332 (2009)
5. Stark, J. Observation of the Separation of Spectral Lines by an Electric Field. *nature* **92**, 401 (1913)
6. Lo Surdo, A. L'analogo elettrico del fenomeno di Zeeman e la costituzione dell'atomo. *L'elettrotecnica* **1**, 624-634 (1914)
7. Foukal, P., Hinata, S. Electric Fields in the Solar Atmosphere – a Review. *Solar Physics* **132**, 307-334 (1991)
8. Foukal, P., Hoyt, C., Gilliam, L. Electric Fields and Plasma Structure in Coronal Magnetic Loops. *Astrophys. J.* **303**, 861-876 (1986)
9. Foukal, P. V., Behr, B. B. Measurements of Electric Fields in Coronal Magnetic Structures. *Proceedings of Solar Coronal Structures*, 177-184 (1994)
10. Foukal, P. V., Behr, B. B. Testing MHD Models of Prominences and Flares with Observations of Solar Plasma Electric Fields. *Solar Physics* **156**, 293-314 (1995)
11. Moran, T., Foukal, P. An Electrograph for Measurement of Macroscopic Electric Fields in Prominences and Flares. *Solar Physics* **135**, 179-191 (1991)
12. Landi Degl'Innocenti, E., Landolfi, M. *POLARIZATION IN SPECTRAL LINES* (Kluwer Academic Publishers, Dordrecht, 2004)
13. Landi Degl'Innocenti, E. The determination of vector magnetic fields in prominences from the observations of the Stokes profiles in the D3 line of Helium. *Solar Physics* **79**, 291-322 (1982)
14. Casini, R. Resonance scattering formalism for the hydrogen lines in the presence of magnetic and electric fields. *Physical Review A* **71**, 062505 (2005)
15. Favati, B., Landi Degl'Innocenti, E., Landolfi, M. Resonance scattering of Lyman-alpha in the presence of an electrostatic field. *Astron. Astrophys.* **179**, 329-338 (1987)
16. Casini, R., Manso Sainz, R. Scattering polarization of hydrogen lines in the presence of turbulent electric fields. *Journal of Physics B* **39**, 3241-3253 (2006)
17. López Ariste, A. et al. Full Stokes Spectropolarimetry of H$\alpha$ in Prominences. *Astrophys. J.* **621**, 145-148 (2005)
18. Rimmele, T. R. et al. The Daniel K. Inouye Solar Telescope – Observatory Overview. *Solar Physics* **295**, 172 (2020)
19. de Wijn, A. G. et al. The Visible Spectro-Polarimeter of the Daniel K. Inouye Solar Telescope. *Solar Physics* **297**, 22 (2022)
20. Wöger, F. et al. The Daniel K. Inouye Solar Telescope (DKIST)/Visible Broadband Imager (VBI), *Solar Physics* **296**, 145 (2021)

21. Krikova, K., Pereira, T. M. D., Rouppe van der Voort, L. H. M. Formation of Hε in the solar atmosphere. *Astron. Astrophys.* **677**, A52 (2023)
22. Mitchell, W. M. Recent Solar Observations at Haveford. *Astrophys. J.* **30**, 75-85 (1909)
23. Ellerman, F. Solar Hydrogen "bombs". *Astrophys. J.* **46**, 298-300 (1917)
24. Rust, D. M. Chromospheric Explosions and Satelite Sunspots. *IAU Symposium* **35**, 77-84 (1968)
25. Kitai, R. On the mass motions and the atmospheric states of moustaches. *Solar Physics* **87**, 135-154 (1983)
26. Georgoulis, M. K., Rust, D. M., Bernasconi, P. N., Schmieder, B. Statistics, Morphology, and Energetics of Ellerman Bombs. *Astrophys. J.* **575**, 506-528 (2002)
27. Isobe, H., Tripathi, D., Archontis, V. Ellerman Bombs and Jets Associated with Resistive Flux Emergence. *Astrophys. J.* **657**, 53-56 (2007)
28. Illing, R. M. E., Landman, D. A., Mickey, D. L. Broad-band circular polarization of sunspots; spectral dependence and theory. *Astron. Astrophys.* **41**, 183-185 (1975)
29. Makita, M. An Interpretation of the Broad Band Circular Polarization of Sunspots. *Solar Physics* **106**, 269-286 (1986)
30. Ruiz Cobo, B. et al. DeSIRe: Departure coefficient aided Stokes Inversion based on Response functions. *Astron. Astrophys.* **660**, 37 (2022)
31. Lehmann, J. C. Nuclear Orientation of Cadmium$^{111}$ by Optical Pumping with the Resonance Line $5^1S_0 - 5^1P_1$. *Physical Review*, **178**, 153-160 (1969)
32. Kemp, J. C., Macek, J. H., Nehring, F. W. Induced atomic orientation, an efficient mechanism for magnetic circular polarization. *Astrophys. J.* **278**, 863-873 (1984)
33. Brooks, D. H., Kurokawa, H., Berger, T. E. An H*α* Surge Provoked by Moving Magnetic Features near an Emerging Flux Region. *Astrophys. J.* **656**, 1197-1207 (2007)
34. Leake, J. E., Lukin, V. S., Linton, M. G., Meier, E. T. Multi-fluid Simulations of Chromospheric Magnetic Reconnection in a Weakly Ionized Reacting Plasma. *Astrophys. J.* **760**, 109 (2012)
35. Khomenko, E., Collados, M., Díaz, A., Vitas, N. Fluid description of multi-component solar partially ionized plasma. *Physics of Plasmas* **21**, 092901 (2014)
36. Lazarian, A., Vishniac, E. T. Reconnection in a Weakly Stochastic Field. *Astrophys. J.* **517**, 700-718 (1999)
37. Shibata, K., Tanuma, S. Plasmoid-induced-reconnection and fractal reconnection. *Earth, Planes and Space* **53**, 473-482 (2001)
38. Bhattacharjee, A., Huang, Y.-M., Yang, H. & Rogers, B. Fast reconnection in high-Lundquist-number plasmas due to secondary tearing instabilities. *Phys. Plasmas* **16**, 112102 (2009)
39. Samtaney, R., Loureiro, N., Uzdensky, D., Schekochihin, A. & Cowley, S. C. Formation of plasmoid chains in magnetic reconnection. *Phys. Rev. Lett.* **103**, 105004 (2009)

40. Loureiro, N. F., Samtaney, R., Schekochihin, A. A., Uzdensky, D. A. Magnetic reconnection and stochastic plasmoid chains in high-Lundquist-number plasmas. *Physics of Plasmas* **19**, 042303-042303-5 (2012)
41. Watanabe, H., Vissers, G., Kitai, R., Rouppe van der Voort, L., Rutten, R. J. Ellerman Bombs at High Resolution. I. Morphological Evidence for Photospheric Reconnection. *Astrophys. J.* **736**, 71 (2011)
42. Reardon, K. The Effects of Atmospheric Dispersion on High-Resolution Solar Spectroscopy. *Solar Physics* **239**, 503-517 (2006)
43. Vissers, G. J. M., Rouppe van der Voort, L. H. M., Rutten, R. J. Automating Ellerman bomb detection in ultraviolet continua. *Astron. Astrophys.* **626**, 4 (2019)
44. Lemen, J. R. et al. The Atmospheric Imaging Assembly (AIA) on the Solar Dynamics Observatory (SDO). *Solar Physics* **275**, 17-40 (2012)
45. Pesnell, W. D., Thompson, B. J., Chamberlin, P. C. The Solar Dynamics Observatory (SDO). *Solar Physics* **275**, 3-15 (2012)
46. Bobra, M. G. et al. The Helioseismic and Magnetic Imager (HMI) Vector Magnetic Field Pipeline: SHARPs – Space-Weather HMI Active Region Patches. *Solar Physics* **289**, 3549-3578 (2014)
47. Metcalf, T. R. Resolving the 180-degree ambiguity in vector magnetic field measurements: The 'minimum' energy solution. *Solar Physics* **155**, 235-242 (1994)
48. Leka, K. D. et al. Resolving the 180° Ambiguity in Solar Vector Magnetic Field Data: Evaluating the Effects of Noise, Spatial Resolution, and Method Assumptions. *Solar Physics* **260**, 83-108 (2009)
49. Gingerich, O., Noyes, R. W., Kalkofen, W., Cuny, Y. The Harvard-Smithsonian reference atmosphere. *Solar Physics* **18**, 347-365 (1971)
50. Gudiksen, B. V. et al. The stellar atmosphere simulation code Bifrost. Code description and validation. *Astron. Astrophys.* **531**, 154 (2011)
51. Carlsson, M. et al. A publicly available simulation of an enhanced network region of the Sun. *Astron. Astrophys.* **585**, 4 (2016)
52. Hansteen, V. H. et al. Bombs and Flares at the Surface and Lower Atmosphere of the Sun. *Astrophys. J.* **839**, 22 (2017)
53. Kawabata, Y. et al. Multiline Stokes Synthesis of Ellerman Bombs: Obtaining Seamless Information from Photosphere to Chromosphere. *Astrophys. J.* **960**, 26 (2024)
54. Uitenbroek, H. Multilevel Radiative Transfer with Partial Frequency Redistribution. *Astrophys. J.* **557**, 389-398 (2001)
55. Anan, T., Casini, R., Ichimoto, K. Diagnosis of Magnetic and Electric Fields of Chromospheric Jets through Spectropolarimetric Observations of H I Paschen Lines. *Astrophys. J.* **786**, 94 (2014)
56. Cox, A. N. Allen's Astrophysical Quantities (Springer Science+Business Media New York, New York, 2002)

57. Trujillo Bueno, J., Štěpán, J., Casini, R. The Hanle Effect of the Hydrogen Ly$\alpha$ Line for Probing the Magnetism of the Solar Transition Region. *Astrophys. J. Lett.* **738**, 11 (2011)
58. Anan, T. Magnetic diffusion in Solar atmosphere produces measurable electric fields [Data set]. Zenodo. DOI 10.5281/zenodo.13709863 https://doi.org/10.5281/zenodo.13709863 (2024)
59. Anan, T. Magnetic diffusion in Solar atmosphere produces measurable electric fields (v1.0.0). Zenodo. DOI 10.5281/zenodo.13621770 https://doi.org/10.5281/zenodo.13621770 (2024)

**Acknowledgements** The research reported herein is based in part on data collected with the Daniel K. Inouye Solar Telescope (DKIST), a facility of the National Solar Observatory (NSO). The authors are grateful to all the staff of DKIST. The NSO is managed by the Association of Universities for Research in Astronomy, Inc., and funded by the National Science Foundation. Any opinions, findings, and conclusions or recommendations expressed in this publication are those of the authors and do not necessarily reflect the views of the National Science Foundation or the Association of Universities for Research in Astronomy, Inc. DKIST is located on land of spiritual and cultural significance to Native Hawaiian people. The use of this important site to further scientific knowledge is done with appreciation and respect. This work was also supported by JSPS Core-to-Core Program. Part of this research is based upon work supported by the National Center for Atmospheric Research, which is a major facility sponsored by the National Science Foundation under Cooperative Agreement No. 1852977 (R.C.). S.K.T.'s work was supported by the National Science Foundation under Grant No. (2307505). The authors thank Dr. P. V. Foukal for helpful discussions about the subject of this study, and acknowledge his pioneering work on the detection of electric fields in solar plasmas. We also thank Dr. V. H. Hansteen and Dr. Y. Kawabata for supporting analysis of a Bifrost simulation result.

**Author Contributions** T.A. analyzed the ViSP data, wrote the manuscript, and led the discussion. A.T., F.W., and T.R.R. lead the DKIST operation team. S.A.J. took the DKIST data. T.A.S. contributes to reducing the DKIST data. T.A. performed DeSIRe inversions with contributions from H.U. and H.S. R.C. synthesized Stokes spectra with electric fields. R.C. and T.A. determined the electric field. K.I., S.K.T., J.W.R., Y.K., A.A., J.Q., and K.R. participated in the data interpretation and made important contributions to the overall science case. All authors commented on the manuscript.

**Competing Interests**

The authors declare no competing interests.

**Author information** Reprints and permissions information is available at www.nature.com/reprints. The authors declare no competing financial interests. Corresponding and requests for materials should be addressed to T.A. (tanan@nso.edu). Some authors including the corresponding author works in the location where the research is conducted using the Daniel K. Inouye Solar Telescope (DKIST).

## Supplementary Notes

**Stokes-*I* profile in the H$\varepsilon$ line of the Ellerman bomb**

The H$\varepsilon$ line at Ellerman bombs has been observed as an absorption at the line center against its broad emission[6]. Supplementary Fig. 4 shows the Stokes-*I* profile of the Ellerman bomb that we present and a quiet region around the Ellerman bomb in the 397 nm spectral range. At the Ellerman bomb, the red wing was more enhanced than the blue wing, while the Ca II blue wing became brighter than that in the quiet region by about 1.23 times. The extra enhancement in the red wing can be the broad H$\varepsilon$ emission due to temperature enhancement in the lower solar atmosphere[6,7].

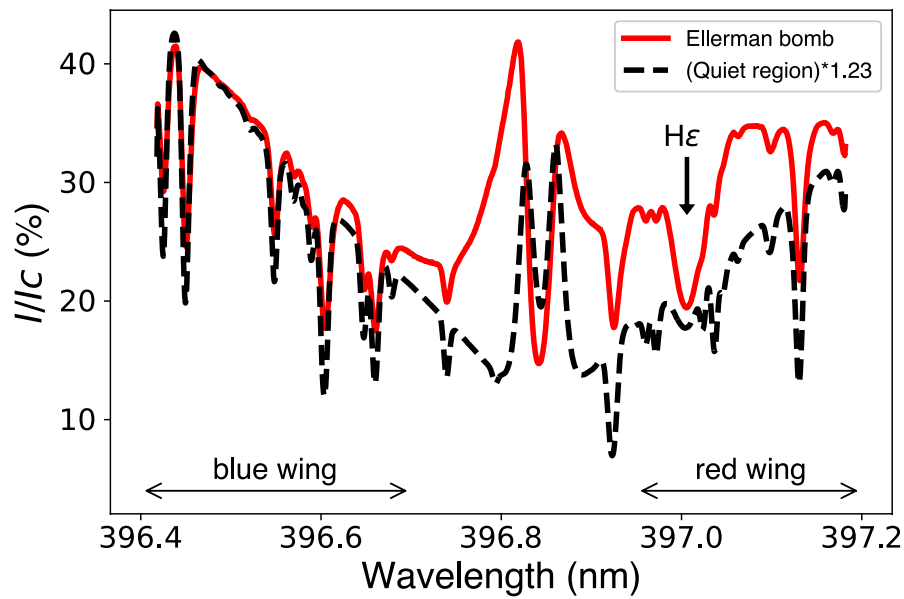

Supplementary Fig. 4 | **Stokes-*I* profiles in 397 nm** of (red) the Ellerman bomb and (black) a quiet region. The intensity for the quiet region was multiplied by a factor of 1.23 to compare the profiles in the Ca II H blue and red wings. Source data are provided as a Source Data file.

**Line-of-sight magnetic field in the photosphere**

Supplementary Fig. 5 shows the line-of-sight and the plane-of-sky components of the magnetic field in the photosphere inferred from the Fe I 630.15 nm line through the Weak Field Approximation method[8]. The Ellerman bomb occurred at the boundary between regions with magnetic fields of opposite directions in the line-of-sight as well.

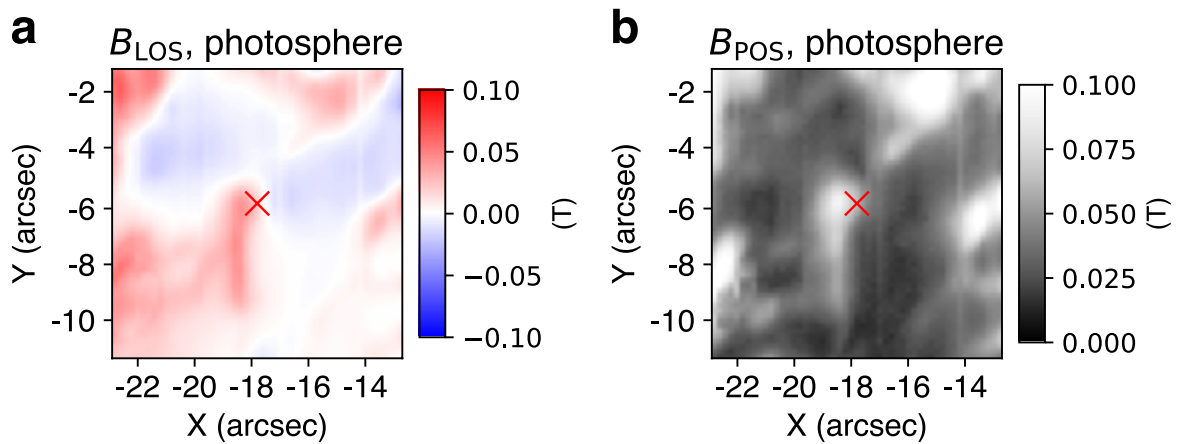

Supplementary Fig. 5| **Magnetic field maps in the line-of-sight frame. a** The line-of-sight component, $B_{LOS}$, and **b** the plane-of-sky component, $B_{POS}$, are shown. The unit for the magnetic fields is tesla, T. The red cross points to the Ellerman bomb and the place where Stokes profiles shown in Figs. 3 and 4 were measured.

**Another Ellerman bomb with electric field signals**

We found another Ellerman bomb that has a symmetric Stokes-*V* profile in the Hε line. The Ellerman bomb, which was bright in the Ca II H line wing (Supplementary Fig. 6a) but not the Ca II H line core (Supplementary Fig. 6b), was observed at 19:40 UT. The Stokes-*I* and *V* profiles are shown in Supplementary Fig. 6 (c, d, and e). The Hε line absorbed the Ca II H line wing intensity between 396.98 nm and 397.02 nm. In the wavelength range, the Stokes-*V/I* profile was mostly negative and symmetric with respect to the line center wavelength (the vertical dashed line). An electric field is required to be taking account to explain this Stokes profile as shown in Fig. 4 (c).

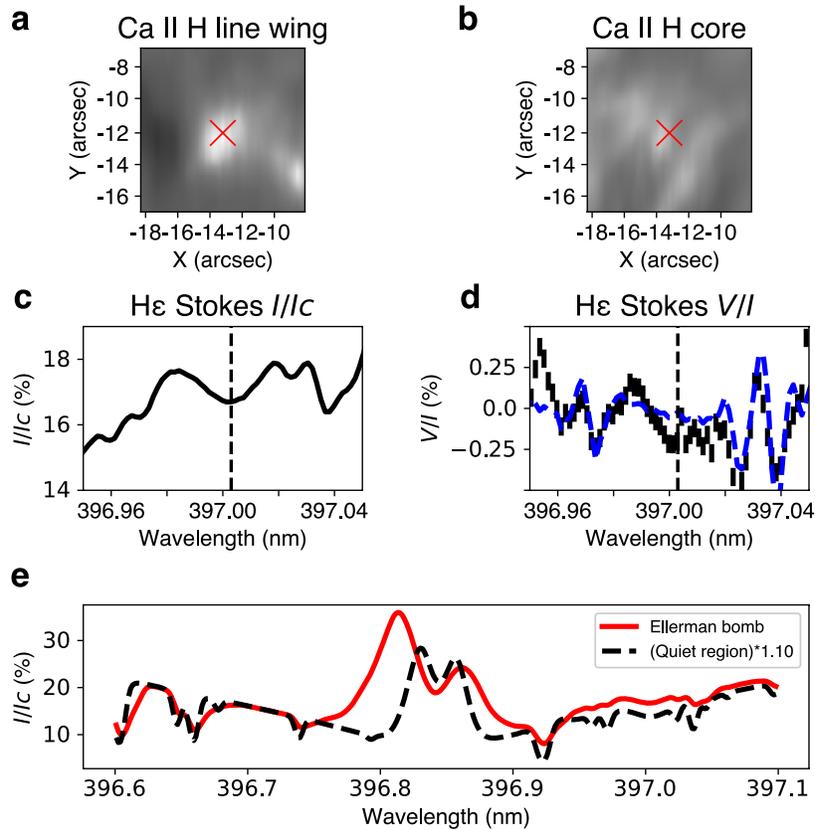

Supplementary Fig. 6 | **Another Ellerman bomb in the Hε line. a** Upper photosphere and lower chromosphere shown by a Visible-Spectro-Polarimeter (ViSP[9]) intensity map in the Ca II H line wing at $-0.05$ nm from the line core. **b** Chromosphere shown by an intensity map at the line core of Ca II H. The red cross points to the Ellerman bomb and the place where Stokes-*I* (**c**) and Stokes-*V/I* (**d**) in the Hε line were measured. Black curves and bars are measured data points. Error bars are shown as the standard deviation in the continuum only for Stokes-*V/I*, because they are too small for Stokes-*I*. The dashed vertical line marks the line center of the Hε line. A blue dashed curve in panel **d** shows an average profile of those pixels that have similar Stokes-*V/I* signals to the Ellerman bomb in weak photospheric lines, i.e., Fe I at 396.96 nm, Cr I at 396.97 nm, Fe I at 397.02 nm, and Ni I at 397.04 nm, from quiet regions around the Ellerman bomb. **e** The Stokes-*I* profiles in a wider spectral range of (red) the Ellerman bomb and (black) a quiet region. The intensity for the quiet region was multiplied by a factor of 1.10 to compare the profiles in the Ca II H blue and red wings. Source data are provided as a Source Data file.

## Supplementary References

1. Lemen, J. R. et al. The Atmospheric Imaging Assembly (AIA) on the Solar Dynamics Observatory (SDO). *Solar Physics* **275**, 17-40 (2012)

2. Pesnell, W. D., Thompson, B. J., Chamberlin, P. C. The Solar Dynamics Observatory (SDO). *Solar Physics* **275**, 3-15 (2012)
3. Ruiz Cobo, B. et al. DeSIRe: Departure coefficient aided Stokes Inversion based on Response functions. *Astron. Astrophys.* **660**, 37 (2022)
4. Hansteen, V. H. et al. Bombs and Flares at the Surface and Lower Atmosphere of the Sun. *Astrophys. J.* **839**, 22 (2017)
5. Kawabata, Y. et al. Multiline Stokes Synthesis of Ellerman Bombs: Obtaining Seamless Information from Photosphere to Chromosphere. *Astrophys. J.* **960**, 26 (2024)
6. Rouppe van der Voort, H. M., Joshi, J., Krikova, K. Observations of magnetic reconnection in the deep solar atmosphere in the Hε line. *Astron. Astrophys.* **683**, 190 (2024)
7. Krikova, K., Pereira, T. M. D., Rouppe van der Voort, L. H. M. Formation of H$\varepsilon$ in the solar atmosphere. *Astron. Astrophys.* **677**, A52 (2023)
8. Landi Degl'Innocenti, E., Landolfi, M. *POLARIZATION IN SPECTRAL LINES* (Kluwer Academic Publishers, Dordrecht, 2004)
9. de Wijn, A. G. et al. The Visible Spectro-Polarimeter of the Daniel K. Inouye Solar Telescope. *Solar Physics* **297**, 22 (2022)